\documentstyle[times,twocolumn,latex8,epsf]{article}
\begin{document}

\title{Effective Monte Carlo simulation on System-V massively parallel
associative string processing architecture}

\author{G\'eza \'Odor\\
Research Institute for Techn. Physics and Materials Science\\
P.O.Box 49 H-1525 Budapest, Hungary\\
odor@mfa.kfki.hu\\
\and
Argy Krikelis\\
ASPEX Microsystems\\
Brunel University, Uxbridge\\
Middlesex, United Kingdom, UB8 3PH\\
Argy.Krikelis@aspex.co.uk
\and
Gy\"orgy Vesztergombi\\
Research Institute for Particle Physics\\
P.O.Box 49 H-1525 Budapest, Hungary\\
veszter@rmki.kfki.hu\\
\and
Francois Rohrbach\\
CERN\\Gen\`eve 23, CH-1211, Switzerland\\F.Rohrbach@cern.ch
}

\onecolumn
\maketitle
\thispagestyle{empty}

\begin{abstract}
 We show that the latest version of massively parallel processing
 associative string processing architecture (System-V) is applicable
 for fast Monte Carlo simulation if an effective on-processor random
 number generator is implemented. Our lagged Fibonacci generator can
 produce $10^8$ random numbers on a processor string of 12K PE-s.
 The time dependent Monte Carlo algorithm of the one-dimensional
 non-equilibrium kinetic Ising model performs 80 faster than the
 corresponding serial algorithm on a 300 MHz UltraSparc.
\end{abstract}

\newpage

\twocolumn
\section{Introduction}

Massively parallelism appears nowadays mainly on the level of MIMD processor
clusters owing to the commercially available cheap building elements
with ever increasing clock speeds. However everybody knows that CPU clock
speed can not increased without limit, and the memory access speeds are
well behind. Therefore redesigning of the high performing architectures are
necessary time to time. One such a direction is the intelligent memory 
(IRAM \cite{iram}) or processor on memory projects. By putting more and more fast 
memory on the silicon surface of the processors (cashes) or processors at the
edges of the memory matrices one can avoid huge (1000 times magnitude) losses
on the connection hardware, buses.

The Massively Parallel Processing Collaboration \cite{aspfin} started a research 
and development of conceptually similar architectures in the early nineties with 
a target of processing large quantities of parallel data on-line. The basic 
architecture was a low level MIMD high level SIMD to fit the best to the requirements.
While the development has stopped with prototype (ASTRA-2) in the physics 
research development collaboration, the founding engineering company ASPEX
continued developing the ASP architecture to produce a special "co-processor" 
for workstations that enhances image processing capabilities.

The latest architecture System-V has already proven its image processing power
\cite{image}. We demonstrate is this work that it is also applicable for
Monte Carlo simulations in statistical physics. In section \ref{sec:SYS} we 
introduce the basics of the hardware of the new architecture, while in sections 
\ref{sec:RG} and \ref{sec:MC} we show how effective random generation and simulations
can be coded. In section \ref{sec:NEKIM} we introduce the statistical physical models
and the time dependent algorithms \ref{sec:DIN} to measure critical exponents. 
More detailed analysis of the results will be discussed elsewhere 
\cite{tobe,odunp}.

\section{The System-V architecture}\label{sec:SYS}

System-V is a specific VMEbus system implementation of the Modular MPC 
(Massively Parallel Computing) architecture. It provides programmable, 
cost-effective yet flexible solutions for high performance data-intensive 
computing applications, such as signal and image processing and 2D and 3D 
image generation problems. The architecture of System-v is modular, so 
that the configuration of processors, memory, data channels and I/O units 
in a system can be independently scaled and exactly balanced to give the 
most economical and effective application solution. Application development 
is achieved by programming in C.

System-V derives its computational flexibility and high-performance from a 
programmable SIMD architecture, namely the ASP 
(Associative String Processor) \cite{Mike}. 
In addition to the SIMD core, high performance RISC processors and custom 
ASICs are used to issue program instruction and manage data I/O.
\begin{figure}[h]
\epsfxsize=80mm
\centerline{\epsffile{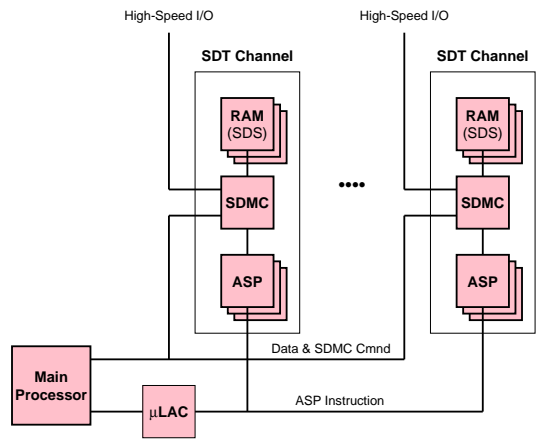}}
\vspace*{0.5cm}
\caption{System-V architecture schematic diagram}
\label{fig1}
\end{figure}

As indicated in Figure \ref{fig1}, the key features of the System-V architecture are 
multiple data channels with overlapped data transfer and processing, independent 
scalability of processors and memory in each channel, and multiple external I/O paths.
The major building block of the architecture is the Secondary Data Transfer (SDT) 
Channel, comprising the ASP, Secondary Data Movement Controller (SDMC) and 
Secondary Data Store (SDS).  SDT channels can be replicated to increase 
SDT bandwidth.  Each channel also has  its own high-speed Tertiary Data Interface 
(TDI) for direct high-speed external I/O.

As described in a following sub-section, the ASP contains an array of Associative 
Processing Elements (APEs) and a Primary Data Store (PDS).  Each APE has a private 
Primary Data Transfer (PDT) channel to the PDS.  In addition, all APEs are connected 
via a flexible communications network.

The PDS is connected to the SDS by the SDMC, which performs secondary data 
transfers between the two.  Its also controls access to the SDS for Tertiary 
Data Transfers (TDT) from external devices or the Main Processor.

The Low-level ASP Controller (uLAC) performs global control of the ASPs. 
The Main Processor can be a  conventional workstation, PC, CPU card or 
micro-processor.  To control System-V it must perform the tasks of Instruction 
Stream Management (ISM) and Data Stream Management (DSM). It does this by issuing 
commands to the uLAC and the SDMCs.
\begin{figure}[h]
\epsfxsize=80mm
\centerline{\epsffile{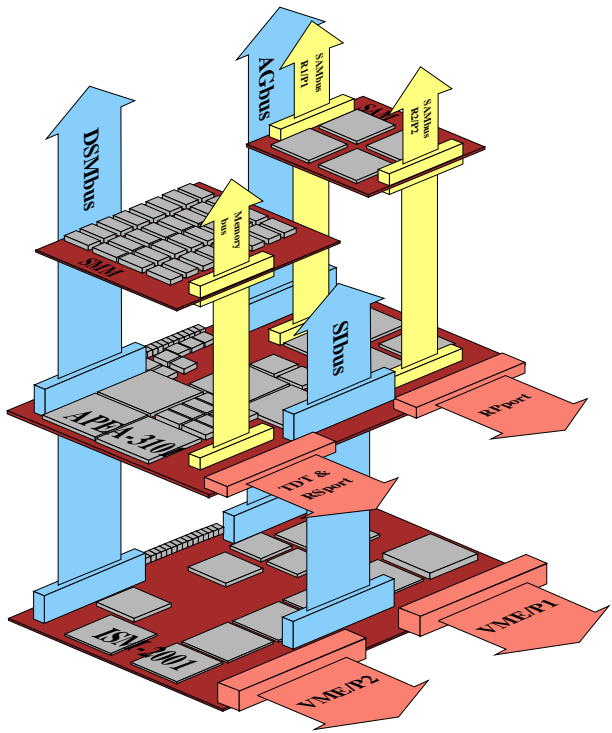}}
\vspace*{0.5cm}
\caption{System-V stackable implementation diagram}
\label{fig2}
\end{figure}

As depicted in Figure \ref{fig2}, comprises a VMEbus motherboard on which sits 
a stack of daughter-boards with the option of smaller mezzanine cards installed 
on-top of the daughter-boards.

In the heart of System-V flexibility and scalability is its stacking buses. 
Three high performance buses called AGbus, SIbus and DSMbus run through all the 
cards in the System-V stack. A further two nested buses provide APE and memory 
expansion on the SDT Channel cards.

The ASP Global Bus (AGbus), a synchronous bus, carries instructions from the 
instruction stream manager on the stack base card to the ASP devices on the APE 
array daughter-boards.

A single System Interface bus (SIbus) connects all the cards in the System-V stack 
to the management processors on the base card and to the VMEbus.

The Data Stream Manager bus (DSMbus) connects the data stream management processor 
on the base card or on a daughter-board to all the cards in the System-V stack between 
it and the next DSM card above it.  There can be up to eight DSMbuses in a single 
System-V stack, all of which can operate in parallel.

The Stackable APE Module bus (SAMbus) allows the number of APEs to be scaled per 
SDT channel by plugging in more SAMs.

Finally, the memory expansion bus allows the amount of memory per SDT channel to 
be scaled by plugging in Stackable Memory Modules (SMM)s.

Multiple System-V SIMD stacks can be connected together in pipeline or processor 
farm topologies to provide even higher levels of performance.

\subsection{System-V building blocks}

The hardware blocks that can be used in building a System-V configuration are 4 cards. 
The base or ISM card combines the ISM and DSM functionalities of System-V. 
It is a VMEbus card that provides the interface between the SIMD stack and the 
rest of the VME system. It features two SPARC processors and a uLAC-1001 
co-processor for DSM and ISM functions respectively.

The APE Array (APEA) card is SIMD stack daughter-boards that implement a Secondary 
Data channel containing an Associative Processing Element (APE) Array, Secondary 
Data Store (SDS) and a Secondary Data Movement Controller. The number of APEs and 
size of memory can be independently scaled by SAM and SMM mezzanine cards.

The Stackable APE Module (SAM) cards contain VLSI chips implementing APEs and 
conform to a standard mechanical and interface specification.  The number
of APEs in a system can be increased by adding more SAMs. The SAM and SAMbus
standards allow existing systems to be simply upgraded as new generations of 
VLSI chips become available.

The SMM-1016 is a stackable SDS memory expansion module that can sustain a 
120 Mbytes/s access bandwidth.

\subsection{Associative String Processor (ASP)}

\begin{figure}[h]
\epsfxsize=80mm
\centerline{\epsffile{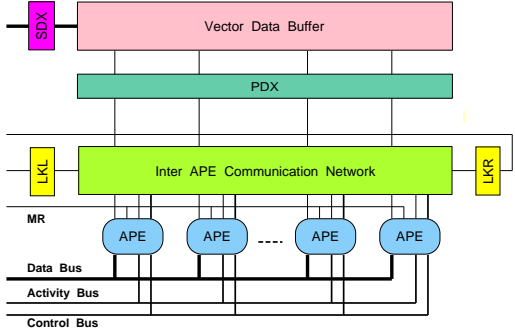}}
\vspace*{0.5cm}
\caption{Associative String Processor architecture}
\label{fig3}
\end{figure}

As mentioned above, the processing core in System-V is an SIMD processing structure 
implemented using ASP Modules. The ASP, shown in Figure \ref{fig3}, is a  modular 
massively parallel and inherently fault-tolerant processing architecture.

At the logical level, the Associative String Processor (ASP) constitutes a 
high-performance cellular string associative processor, whereas, at the 
physical level, the ASP is implemented as a bit-parallel word-parallel 
associative parallel processor.  The ASP is a programmable, homogeneous and 
fault-tolerant fine-grain SIMD massively parallel processor incorporating a 
string of identical Associative Processing Elements (APEs), a reconfigurable 
inter-processor communication network and a Vector Data Buffer for 
fully-overlapped data input-output as indicated in Figure \ref{fig3}.

\begin{figure}[h]
\epsfxsize=80mm
\centerline{\epsffile{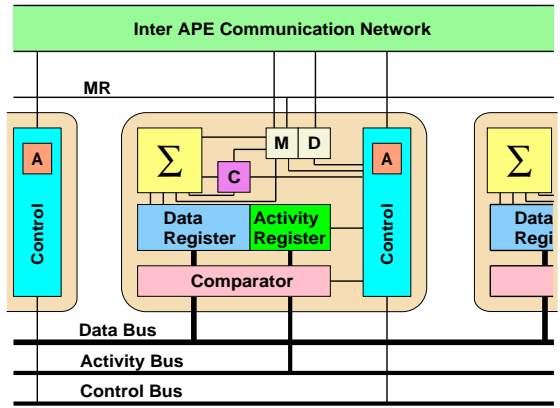}}
\vspace*{0.5cm}
\caption{Associative Processing Element architecture}
\label{fig4}
\end{figure}

Each APE, depicted in Figure \ref{fig4}, incorporates a 64-bit Data Register 
and a 6-bit Activity Register, a 70-bit parallel Comparator, a single-bit 
full-adder, 4 status 
flags and control logic for local processing and communication with other APEs. 
The 6-bit Activity Register is used to select subsets of APEs for subsequent 
parallel processing.

The APEs are connected via the Inter APE Communication Network, which supports 
the navigation of data structures and implements a simply-scalable, fault-tolerant 
and dynamically-reconfigurable (to support emulation of common network topologies) 
tightly-coupled APE interconnection strategy. As an activity-passing, rather than 
a data-passing, network, the Inter APE Communication Network reduces data transfers. 
Time-consuming data transfers are only executed on the Inter APE Communication 
Network, if they cannot be otherwise avoided. Most significantly, the APE 
interconnection strategy supports simple unlimited modular network extension, 
via the LinK Left (LKL) and LinK Right (LKR) ports (shown in Figure \ref{fig3}) 
to enable tailoring of parallel processing power to match user requirements.

The Inter APE Communication Network is derived from a shift register and a 
chordal ring, which supports 2 modes of inter APE communication: 
asynchronous bidirectional single-bit communication, to connect APE sources 
and corresponding APE destinations of high-speed activation signals, implementing 
a fully-connected dynamically-configured (programmer-transparently) permutation, 
and broadcast network for APE selection and inter-APE routing functions; 
synchronous bidirectional multi-bit communication, via a high-speed bit-serial 
shift register for data/message transfer between APE groups. Thus, the 
interconnection strategy adopted for the ASP supports a high degree of parallelism 
for local communication and progressively lower degrees of parallelism for longer 
distance communication. In particular, the chordal ring topology enables the Inter 
APE Communication Network to be implemented as a hierarchy of APE groups. 
Thus, communication times are significantly reduced through automatic bypassing 
of those groups that do not include destination APEs. In a similar way, namely 
through bypassing of faulty groups of APEs, fault tolerance of the ASP architecture 
is guaranteed.

In operation, data are distributed over the APEs and stored in the 
local Data Registers. Successive computational tasks are performed on the stored
data and the results are dumped via the PDX, to the Vector Data Buffer (shown
in Figure \ref{fig3}). The ASP supports a form of set processing, in which a subset of 
active APEs (i.e., those which match broadcast scalar data and activity values) 
support scalar-vector (i.e., between a scalar and Data Registers) and vector-vector 
(i.e., within Data Registers) operations. Matching APEs are either directly activated 
or source inter APE communications to indirectly activate other APEs via the Inter APE 
Communication Network. The Match Reply (MR) line to the control interface provides 
feedback on whether none or some APEs match. The APE can operate in three different 
data modes dictated by the Data Register configuration. The supported modes are: 
storage and bit-parallel processing of two 32-bit words or four 8-bit byte fields 
and storage and bit-serial processing of one to three bit-fields of varying length 
(of no more than 64 bits per field). The instruction set is based on 4 basic 
operations, match, add, read and write. In order to achieve bit-level masking, 
during match and write operations, the corresponding byte and bit fields of the 
Data Bus are represented with ternary (i.e., 2-bit) digits.

\section{The random number generator}\label{sec:RG}

The random number generator we used here is a lagged Fibonacci, algorithmically
the same that was described in \cite{EM97} in more details.
We exploit the orthogonal manipulation capabilities of the ASP string and 
generate 160 bit long pseudo random numbers along the string with fast 
"look ahead carry" operations.
\begin{equation}
x_i =  x_{i-17} \pm x_{i-5} \pm c  \label{Fib}
\end{equation}
We segment the ASP string to $160$ APE parts and so we generate
No. of APE-s / 160 such numbers in parallel. 
We take one bit of the generated numbers for each processor as a bit of
a 18-bit integer random number. hence to build up a full 18-bit random number for
an APE we have to repeat the $x_i$ generation step 18 times. The lag columns 
are moving in a circular way in the 18-bit memory field of APE-s.
The carries of eq. \ref{Fib} are transmitted back between two addition steps
to the beginning of the segments by the effective "Activity Link" operation.
So we can generate a 18-bit random number -- which is a compromise of the
on-processor memory and necessary resolution -- within a few clock cycles in 
all APE-s independently of the system size.

These integer random numbers can be thresholded in parallel by constants 
($p*2^{18}$) that are stored in the memory of the ISM . By this operation we 
can tag processors with probability $p$.
Practically we could achieve a $\sim 10^4$ update/ sec. that means $\sim 10^8$
random numbers in every second if the system size is 12K.

Testing of the random number generator was through comparing the simulation 
results with results obtained on a serial computer.

\section{Time dependent Monte Carlo simulation}\label{sec:DIN}

Time dependent Monte Carlo simulation suggested by \cite{Gras79} has become a 
very precise and effective tool in statistical physics.
We start the system from a state that is usually random or near to the
absorbing state and follow the evolution of its statistical properties 
for a few thousand time steps. 

In general one usually "measure" the following quantities
\begin{itemize}
\item survival probability $p_s(t)$ of the initial seed
\item order parameter density $\rho(t)$
\end{itemize}

The evolution runs are averaged over $N_s$ independent runs
for each different value of $p$ in the vicinity of $p_c$.
At the critical point we expect these quantities to behave
in accordance with the power law as $t\to\infty$, i.e.
\begin{equation}
p_s(t)\propto t^{-\delta} \ , \label{statd}
\end{equation}
\begin{equation}
\rho(t)\propto t^{\alpha} \ , \label{state}
\end{equation}

For estimating critical exponents and the critical point there is a 
very effect way by analysing the local slopes of the $\log - \log$
quantities. Example for $\rho$
\begin{equation}
\alpha(t) = {\ln \left[ \rho(t) / \rho(t/m) \right] \over \ln(m)}
\end{equation}
where we use $m=8$ usually. 
In the case of power-law behaviour we expect $\alpha(t)$ to be straight 
line as $1/t \to 0$, when $p = p_c$. The off-critical
curves should possess curvature. Curves corresponding to $p > p_c$
should veer upward, curves with $p < p_c$ should veer down.

\section{The NEKIM model}\label{sec:NEKIM}

The research of phase transitions of non-equilibrium models is in the forefront 
of statistical physics. Very few models are solved and the universality class 
picture of equilibrium systems can not be directly transferred to here.
The lack of detailed balance condition
\begin{equation}
P(\{s\}) W(\{s\}\to\{s^,\}) = P(\{s^,\}) W(\{s^,\}\to\{s\})
\end{equation}
--- where $P(\{s\})$ is the probability of a state, and $W$ is the transition 
probability  --- enables arbitrary noise and this seems to have an affect on the
ordered state influencing the critical scaling behaviour. 
This suggests much richer behaviour than in equilibrium statistical systems.
Contrary to this for a long time there has been only one phase transition 
universality class known according to which models are categorised 
in equilibrium systems. Namely every continuous phase transition to an absorbing
state have ultimately been found to belong to the class of Directed Percolation
or Reggeon Filed theory \cite{Cardy}. Theoretical investigations have shown the 
robustness of this class \cite{DP1,DP2}. There are a few exceptions found up to now.
One such an exceptional class is the Parity Conserving (PC) universality class 
in 1d, which was named after that the number of particles is conserved modulo 2.
Later it was realized \cite{Cardy-Uwe} that more precisely the special dynamics 
the "Branching and annihilating random walk with even number of offsprings" 
(BARWe) is responsible for this non-DP behaviour since the underlying field 
theory possesses a special symmetry in this case.
The field theoretical description of this class has not given quantitatively
precise results and we can rely on simulation results for critical exponents and
scaling relations.
We have been investigating one representative of the PC class for some years
namely a special non-equilibrium kinetic Ising model in 1d (NEKIM) 
\cite{Racz,men94,meor,meod}.

The Ising model is the simplest system that is capable of describing of 
ferro-magnetism on the basis of collective action of spin variables ($s_i$).
The generalisation of the static model that involves spin flip dynamics was 
done by Glauber \cite{Glau}first. The Glauber model is exactly solvable in 1d 
and the kink variables ($k_i$) ('01'or '10' pairs ) between the ordered domains 
has been found to exhibit annihilating random walk.
Other kind of dynamics can also be introduced that lead to the same equilibrium 
state, for example the spin number conserving Kawasaki dynamics (see \cite{Kawa}).
It was suggested that if we apply different kind of dynamics alternately we can
create a system that does not have an equilibrium state described by Boltzmann
distribution but may possess a steady state in which the some global parameters are
constant (magnetisation example) similarly to the eq. systems but others are not
(example particle or energy currents can flow through). The alternating application
of Glauber spin-flip and Kawasaki spin-exchange was proposed by \cite{Racz} and
it was discovered by \cite{men94} that there is non-temperature driven
phase transition in which the kinks exhibit BARWe dynamics and so the universality
class is PC. Originally this transition has been shown for zero temperature
spin-flip (because in 1d any finite temperature causes such a strong fluctuations
that destroy the order) plus an infinite temperature (process that does
not depend on the local neighbourhood of spins) spin-exchange. 
The spin exchanges don't destroy the order because they don't do anything inside
the ordered domains.

In this work we investigate numerically the generalisation of this model for 
finite temperature spin-exchange ($T_K$) and investigate its effect on the
transition \cite{tobe}. We define the model on the 1d ring (i.e. periodic 
boundary conditions) with the transition probabilities:
\begin{itemize}
\item kink random walk : $w_W = \Gamma(1-\delta)$ 
\item kink annihilation : $w_A = {\Gamma\over 2}(1+\delta)$
\item spin-exchange : 
$w_{i,i+1}={p_{ex}\over 2}(1-s_is_{i+1})(1-{\gamma\over 2}
(s_{i-1}s_i+s_{i+1}s_{i+2})$
\end{itemize}
where the parameters we used are : $\Gamma=0.35$, $p_T=\exp(-4J/kT_{K})$, 
$\gamma=(1-p_T)/(1+p_T)$, $p_{ex}=0.239$. The free parameters are : $\delta$, $T_{K}$.

We used one site per processor mapping and therefore a parallel updating 
version of the NEKIM had to be invented in order to exploit the resources 
of the ASP. To realize the above processes "two lattice update" was employed 
in case of the spin-flip. That means that in one step the "even" lattice 
sites and in the next step the "odd" lattice sites were updated. In case of 
the subsequent spin-exchange a "three lattice update was performed. 
Further algorithmic details will be presented in the next section.

\begin{figure}[h]
\epsfxsize=80mm
\centerline{\epsffile{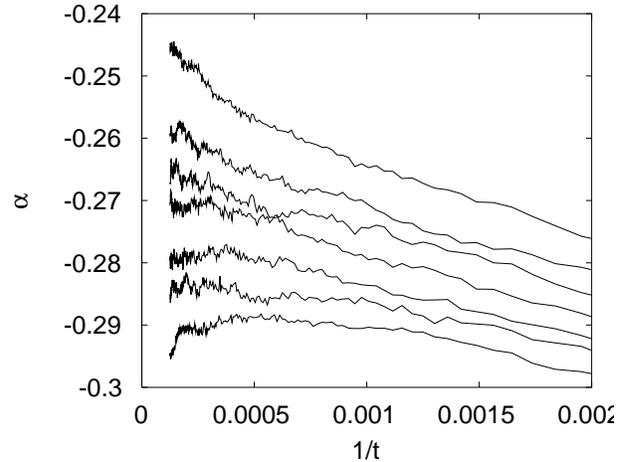}}
\vspace*{0.5cm}
\caption{Local slopes of $\log(\rho_{kink}(t))$ versus $\log(1/t)$ 
in the NEKIM simulations at for 
$p_T=0.26$, $0.265$, $0.27$, $0.275$, $0.28$, $0.285$, $0.29$ 
(from bottom to top curves). 
Scaling can be observed for $p_T=0.27(1)$, with the exponent 
$\alpha\sim 0.28$ in agreement with PC order parameter density decays. }
\label{pT}
\end{figure}

The time dependent simulations have been performed on a $L=3040$ size 
(number of APE-s) ring by starting from random initial states and following the
kink density up to $t_{MAX}=8000$. The number of runs over which the statistical
averaging was performed was $3 - 5\times 10^4$ for each $p_T$ parameter.
First the critical point was located for a fixed value of $\delta=-0.362$, 
$p_{ex}=0.239$ by varying $p_t$. As one can see on the local slope Figure 
\ref{pT} it is about $p_T = 0.27$, because the other curves corresponding to other
$p_T$-s deviate from scaling as $t\to\infty$. We can read off from the ordinate of
the graph that the corresponding critical exponent is $0.28(1)$ that is in a
good agreement with the PC class value \cite{men94}.

\begin{figure}[h]
\epsfxsize=80mm
\centerline{\epsffile{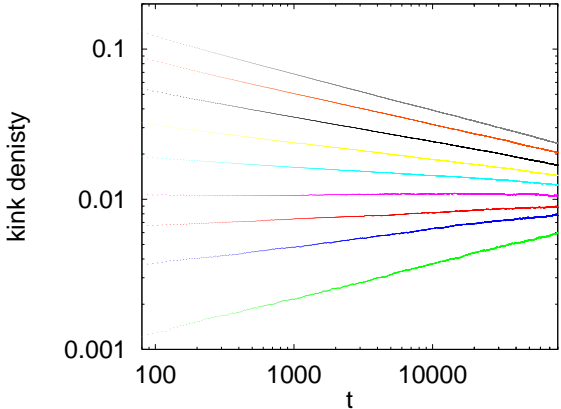}}
\vspace*{0.5cm}
\caption{$\log(\rho_{kink}(t))$ versus $\log(t)$ in the NEKIM simulations at for 
$\sigma=0, 0.1, 0.2 ...,1$ initial conditions (from bottom to top curves). }
\label{rholri}
\end{figure}

We performed timing measurements on a System-V with 12K APEs and with 4 Mbytes 
of memory (that occupy 3 6U VME slots) and is attached to a SparcStation 5/64 
that serves as host for the applications. On that machine $\sim 2 \times 10^{-8}$ 
second is necessary to update a site. 
In comparison the serial simulation Fortran program that was run on a 
DEC-ALPHA 2000 station has achieved $2 \times 10^{-6}$ sec. / site speed..
The other thing that we investigated is the effect of long-range initial 
conditions that has just recently been shown to be relevant \cite{Haye} 
in case of DP transitions.
The initial states with $<k_ik_{i+x}> \ \propto x^{-(1-\sigma)}$ kink-kink 
correlations and even numbered kinks are generated by the same serial algorithm as described 
and numerically tested in ref. \cite{Haye} in the $\sigma\in (0,1)$ interval. 
We required the even-numbered initial kink sector because the kink number is
conserved mod 2.  The spin states are assigned to the kink configuration and 
are loaded to the ASP string time to time during the trial runs. 
The kink density has been measured in $L=12000$ sized systems up to 
$t_{max}=80000$ time steps such that we can observe good quality of scaling 
for three decades in the $(80,80000)$ time interval (see  Figure \ref{rholri}).
As one can see there is an increase with exponent $\alpha \sim 0.28$ in the 
kink density for $\sigma=0$, where in principle only one pair of kinks is 
placed on the lattice in agreement with former simulation results \cite{meod}. 
On the other extreme case for $\sigma=1$ the kink density decays with 
$\alpha\sim -0.28$ exponent again in agreement with our expectations i
\cite{men94}.
In between the exponent $\alpha$ changes continuously as the function of 
$\sigma$ and changes sign at $\sigma=0.5$. That means that the state 
generated with $\sigma=1/2$ is very near to the $t\to\infty$ steady state 
limit \cite{odunp}.

\section{Monte Carlo simulation algorithm}\label{sec:MC}

\begin{figure}
\epsfxsize=80mm
\centerline{\epsffile{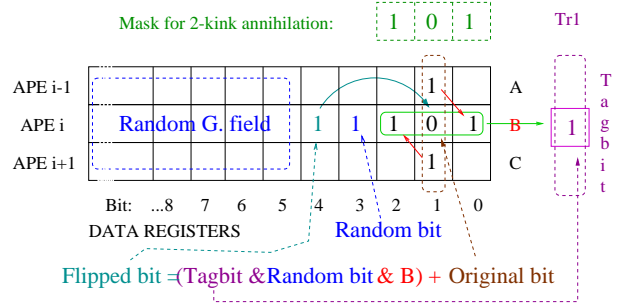}}
\vspace*{0.5cm}
\caption{Data processing in memories of APEs corresponding to a sub-lattice 'B' update.
The kink-s between different spins ('101') are annihilated with probability 
$w_A=\Gamma/2(1+\delta)$ ('111'). First the spins are propagated to the left and
right neighbours, than a global mask ('101') is compared with B tagged APEs
and which are set with probability $w_A$. The spin flipping is done by binary adding
and the result is written back to the original bit.}
\label{kinka}
\end{figure}

\begin{figure}[h]
\epsfxsize=80mm
\centerline{\epsffile{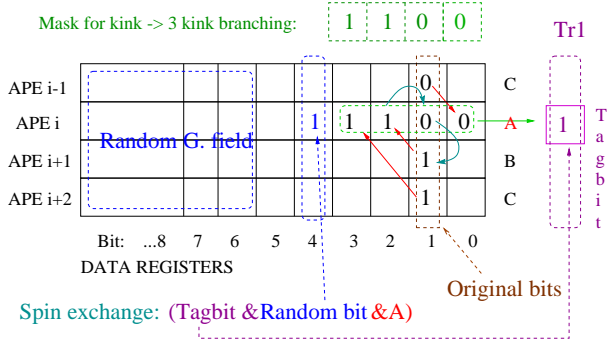}}
\vspace*{0.5cm}
\caption{Data processing in memories of APEs corresponding to a sub-lattice 'A' update.
The kink offsprings are created with probability $w_C=4 p_{ex} p_T /(1+p_T)$. 
First the spins are propagated to the left and right neighbours, than a global mask 
('1100') is compared with all 'A' tagged APEs and which are set with probability $w_C$. 
The spin exchange is performed by overwriting the ASP Data Register bit column containing
the original state.}
\label{bra}
\end{figure}

The low-level ASP part of the simulation is similar to what was described in
\cite{EM97}. We map the 1d system onto a non-segmented string (left and right
ends of the string are connected) (but in case of random generation steps we 
re-segment it to 160 PE substrings to avoid long communications).
The left and right neighbour informations are shifted simultaneously to each
APE and we apply the transition rules with ternary masks and the random
threshold conditions. Since we update in two (and three) sub-lattices every
second and third APE are marked by Activity bits (A,B,C) that is fixed from the
beginning of the simulations and we take into account these Activity bit
conditions as well when doing spin-flip (by the 'Add' operation) or the spin
exchange. See Figure \ref{kinka} for the APE representation of the kink 
annihilation process. For kink random walk the only difference is that we
use different masks (100 or 110 ... etc.).
In case of the finite temperature spin-exchange we have to take into
account even more conditions (i.e. four spin states) as shown on Fig. \ref{bra}.
The spin state is transformed to kink state within a single (Shift+Add) operation 
cycle and an internal "Global Match reply" mechanism examines if the system evolves
to the kink-free ordered state. If it is in the ordered state the whole time
development cycle stops, because in the absorbing state there are no more changes 
take place. The entire kink state is dumped by the efficient PDX exchange within a
few clock cycles to the PDS where a built in hardware mechanism counts the number 
of '1'-s overlapped in time with the MC updates. We invoke this dumping in every 
8 MC updates only to keep a time balance with the counting. 

The site update time in this case slightly smaller than in case of the finite 
temperature NEKIM : $10^{-8}$ second. The serial version of the program run 
on a SUN station with UltraSparc 300 MHz processors achieved $8\times 10^{-7}$ 
site update speed. That means that the System-V algorithm is 80 faster.
We made a test run on the FUJITSU AP3000 parallel supercomputer with $n=16$ nodes 
as well where the parallelism has been exploited on the level of trial samples. 
We found that the System-V program is a factor of 5 faster than the simulation
on the AP3000.

\section{Conclusions}
In this work we have shown that the System-V image processing architecture is capable
for Monte Carlo simulation with a performance far most exceeding present super-computing
technologies if an effective on-processor random generator is invented.
\begin{table}[h]
\begin{tabular}{|c|c|c|c|c|}
\hline
System-V & ALPHA 2000 & U-Sparc 300 & AP3000 \\
\hline
$10^{-8}$& $2\times10^{-6}$ & $8\times 10^{-7}$ & $5\times 10^{-8}$ \\ 
\hline
\end{tabular}
\caption{\em Site update times in seconds of the NEKIM simulation}
\end{table}
This high performance is possible owing to the data movement minimisation of
the content addressable, associative string processing. That means that inside the
most time consuming core of the algorithm are no processor - memory communications
via buses but each PE acts on its local data. This can be achieved by fine grain 
parallelism: one processor per system site mapping. Of course the technology limits 
the number of PE-s and local memories on a single chip but the modular building 
structure enables to build systems with more and more (even millions in the next 
generations) APEs.

Even if input-output is required during the algorithm (like measurements on the state 
at time to time) the fast and parallel data exchanger PDX does not interrupts the 
internal loops considerably. Other applications like surface growth model simulations 
are in progress and preliminary ratings has already shown similar high performances
\cite{letter,dimer}.

We have implemented one-dimensional system simulations here but we note that since the 
whole architecture is primarily designed for vision and image processing we expect 
that two or more dimensional simulations can also be effectively coded. 
There are built in hardware and software tools for cutting the data to patches and 
loading dumping and rotating them in parallel with the string processing.

Therefore we think that this kind of MPC SIMD modules are good candidates 
for building blocks in large scale, "Grand Challenge" simulation architectures.

\section{Acknowledgements}
We thank N\'ora Menyh\'ard for the stimulating discussions and R. Bishop for the
help in programming System-V. G.\'O. thanks 
ASPEX for the hospitality on the System-V training course and the access
to the machine.

\nocite{iram,aspfin,image,tobe,odunp,Mike,Gras79,EM97,DP1,DP2,Cardy,Cardy-Uwe,
Racz,men94,meor,meod,Glau,Kawa,Haye,letter,dimer}

\bibliographystyle{latex8}
\bibliography{CAsim}

\end{document}